\documentclass[%
	twocolumn,%
        aps,prl,showpacs,%
        amsmath,amssymb]{revtex4}
\usepackage{graphicx}
\usepackage[mathscr]{eucal}
\usepackage{color}                        % Comment out for submission
\definecolor{SLvermilion}{cmyk}{0,.8,1,0} % Comment out for submission

\def\atv5{} % Default: switching off the highlight
\def\atv7{} % Default: switching off the highlight
\def\atv7{\color{SLvermilion}} % Comment out for submission

\newcommand{\pdfrac}[2]{\frac{\partial #1}{\partial #2}}

\newcommand{\sgn}{{\rm sgn}}

\newcommand{\bvec}[1]{\boldsymbol{#1}} % Bold Vector

\def\toexp{\mathop{\rm exp}}
\newcommand{\Texp}{\toexp_{\leftarrow}}

\newcommand{\pauli}{\hat{\sigma}}

\newcommand{\up}{\uparrow}
\newcommand{\down}{\downarrow}

\newcommand{\ket}[1]{|{}#1{}\rangle}
\newcommand{\bracket}[2]{\langle{}#1{}|{}#2{}\rangle}

\newcommand{\AF}{A}
\newcommand{\AD}{A^{\rm D}}
\newcommand{\AntiTexp}{\toexp_{\rightarrow}}

\newcommand{\lambdaR}{\alpha}
\newcommand{\lambdaI}{\beta}

\begin{document}
%%%
%%% Title page
%%%
\date{{\sf DRAFT \today}}
%\date{\today}

\title{Exotic quantum holonomy induced by degeneracy hidden in complex parameter space}
%\title{Riemann sheet structure behind exotic quantum holonomies}
%% From SWK: "exceptional points" might be a jargon
%% From TC: avoid Cheon's XX
%\title{Non-Hermitian Exceptional points behind Cheon's holonomies}

\author{Sang Wook Kim}
\email{swkim0412@pusan.ac.kr}
%%\homepage{}
\affiliation{Department of Physics Education, Pusan National University,
  Busan 609-735, South Korea}

\author{Taksu Cheon}
\email{taksu.cheon@kochi-tech.ac.jp}
%%\homepage{}
\affiliation{Laboratory of Physics, Kochi University of Technology,
  Tosa Yamada, Kochi 782-8502, Japan}

\author{Atushi Tanaka}
\email{tanaka@phys.metro-u.ac.jp}
%%\homepage{http://www.comp.tmu.ac.jp/tanaka_atushi/}
\affiliation{Department of Physics, Tokyo Metropolitan University,
  %%Minami-Osawa,
  Hachioji, Tokyo 192-0397, Japan}

%%% length limit ? (within 600 chars)
\begin{abstract}
An adiabatic change of a bound state along a closed circuit in the parameter space can induces holonomies not only in the phase of the state, but also in the associated eigenspace and eigenvalue. The former is the well-known Berry phase while the latter, namely the exotic holonomy, is found
a decade ago and its origin has not been understood yet. By extending the parameter into the complex number, the correspondence of the exotic holonomies and the degeneracy of the non-Hermitian Hamiltonian, or the exceptional points, is revealed. We show that this explains all the known non-trivial characteristics of the exotic holonomies.
\end{abstract}

%% PACS 2008
%% 03.65.Vf Phases: geometric; dynamic or topological
%% 03.65.Ca Formalism
%%
%% cf. 42.50.Dv Quantum state engineering and measurements (see also
%%          03.65.Ud Entanglement and quantum nonlocality, e.g.,
%%          EPR paradox, Bells inequalities, GHZ states, etc.)
%        03.65.-w Quantum mechanics [see also 03.67.?a Quantum
%                 information; 05.30.?d Quantum statistical mechanics;
%                 31.30.J? Relativistic and quantum electrodynamics (QED)
%                 effects in atoms, molecules, and ions in atomic
%                 physics]
%        02.40.-k Geometry, differential geometry, and topology (see also
%                 section 04 Relativity and gravitation)

\maketitle

An adiabatic change of a quantum system along a closed path $C$ in the parameter space may induce discrepancies. Among them, the phase holonomy~\cite{Berry:PRSLA-392-45,Simon:PRL-51-2167} has been thoroughly investigated. Suppose the system is initially prepared at an eigenstate $\ket{\xi_n}$ and evolved adiabatically along $C$, where spectral degeneracy is assumed to be absent. If the dynamical phase is eliminated from the adiabatic time evolution, the finial state coincides with $\ket{\xi_n(C)}$ that is obtained by the parallel transport from $\ket{\xi_n}$ along $C$~\cite{Simon:PRL-51-2167}. The phase holonomy or the geometric phase resides in the phase of
$\bracket{\xi_n}{\xi_n(C)}$~\cite{endnote:noncyclic}.
In the presence of the spectral degeneracy, Wilczek and Zee pointed out that $\ket{\xi_n(C)}$ and $\ket{\xi_n}$ need not be parallel, where the noncommutative extension of the geometric phase is introduced~\cite{Wilczek:PRL-52-2111}.

{%\bf
A decade ago,} a somewhat new kind of holonomies, referred to as the exotic holonomies, has been reported in Ref~\cite{Cheon:PLA-248-285}. Even if the spectral degeneracy is absent so that no Wilczek-Zee's phase holonomy takes place, $\ket{\xi_n}$ and $\ket{\xi_n(C)}$ are not only different from each other but also found to be orthogonal. In fact, $\ket{\xi_n(C)}$ is parallel with another initial eigenstate, namely $\ket{\xi_{n'}}$ ($n'\ne n$). This is nothing but the so-called {\em eigenspace holonomy}, which is characterized by the matrix $M_{mn}(C)\equiv\bracket{\xi_{m}}{\xi_n(C)}$.
%Namely, the discrepancy resides in the eigenspace.
%To quantify it, we introduce matrix elements
%$M_{mn}(C)\equiv\bracket{\xi_{m}}{\xi_n(C)}$.
The holonomy matrix $M(C)$ describes both the phase and the eigenspace holonomy in a unified way~\cite{Cheon:arXiv:0807.4046}.
%% AT-2009-01-20: Fix the description how the phase holonomy resides in $M(C)$
%It is a signature of the eigenspace holonomy that $M(C)$ has a permutation matrix as a factor, while the reminant factor of $M(C)$ is a diagonal %matrix that contains phase factors to describe the usual phase holonomy.
%%{\bf
It is shown that $M(C)$ is divided into two factors, namely a permutation matrix, which is the signature of the eigenspace holonomy, and a diagonal matrix consisting of phase factors so as to describe the usual phase holonomy.
%%}
%%It is a signature of the eigenspace holonomy that $M(C)$ is a permutation matrix, while the usual phase holonomy is expressed by an identity matrix multiplied by a phase factor.
%%
The eigenspace holonomy gives rise to the corresponding {\em eigenvalue} holonomy since the eigenvalues and the eigenvectors have a one-to-one correspondence.

%% AT-2009-02-13: I'm afraid that "... in a system free from a degeneracy"
%%                implies that it is not mysterious in degenerate systems.
It has been mysterious why and how such exotic holonomies, the eigenspace and the eigenvalue holonomies,
{%\bf
emerge}~\cite{Cheon:PLA-248-285,Cheon:arXiv:0807.4046,%
Tanaka:arXiv:0902.2878,%
Tsutsui01,PRL-98-160407,Miyamoto:PRA-76-042115}.
According to Johansson-Sj\"{o}qvist's theorem, which is a generalization of Longuet-Higgins' theorem~\cite{Herzberg:PRSL-344-147}, the spectral degeneracies should take place in a surface $S$ enclosed by $C$ when a nontrivial phase holonomy exists along $C$~\cite{Johansson:PRL-92-060406}. Furthermore, if $C$ lies near a degeneracy point, the Berry phase is proportional to the solid angle subtended by $C$ at the degeneracy point in the parameter space~\cite{Berry:PRSLA-392-45}. It is natural to expect that the Johansson-Sj\"{o}qvist theorem is also applicable to the exotic holonomies. However, it does not seem to be the case since the parameter space of the minimal model giving rise to the exotic holonomies described by $2\times2$ matrix consists of only a one dimensional loop, i.e. $S^1$, so that it offers no room for the theorem ~\cite{PRL-98-160407,Miyamoto:PRA-76-042115}.

%%{\bf
In order to understand the origin of the exotic holonomy we extend the parameter space into complex regime.
%%}
This allows us to access the {\em hidden} degeneracy of the {\em unphysical} complex eigenvalues, which is known as the exceptional points (EPs) in the context of the non-Hermitian Hamiltonian describing open quantum systems~\cite{KatoExceptionalPoint,Heiss00,Dembowski:PRL-86-787,Dembowski03,Dembowski:PRE-69-056216}.
%More precisely, a path in the complexified parameter space can
%run across a branch cut emanating from an EP.
An EP forms a branch point in a parameter space implying eigenfunctions and eigenvalues are no longer single-valued. We find that the resulting structure of a Riemann surface explains all the key characteristics of the exotic holonomies. It is noted that the Berry phase around an EP is defined in a conventional way along a closed path in the Riemann surface, e.g. encircling the loop {\em twice} in the case that the EP forms a square-root branch point~\cite{Dembowski:PRL-86-787}. It is known that when encircling a closed path including an EP {\em once}, the two eigenstates are exchanged with each other, so do the eigenvalues. Considering the Riemann sheet constructed from a square-root branch point, the holonomy associated with an {\em open} path, namely encircling an EP {\em once}, is relevant to our study. Such an open path induces a permutation with possible sign changes among eigenvectors.

The aim of this letter is to reveal the role of the EPs hidden behind the exotic holonomies by extending the parameter into the complex number. It is shown, in particular, that the non-trivial topology of the parameter space around the EPs essentially governs the holonomy matrix $M(C)$, which substantiates a unified theory of quantum holonomies including the exotic ones introduced in Ref.~\cite{Cheon:arXiv:0807.4046}. It gives us an important message that {\em one might encounter the unexpected non-trivial holonomies induced by the degeneracy hidden in the unphysical domain.}

We explain our idea through an analysis of a periodically kicked spin-$\frac{1}{2}$ system, which offers the simplest example of the exotic holonomies~\cite{PRL-98-160407,Cheon:arXiv:0807.4046}. Note that the applicability of our idea is not limited into this particular case. The Hamiltonian of the model system is given as
\begin{align}
  \label{eq:defHamiltonian}
  \hat{H}(t; \lambda)
  = \mu\hat{P}({\bvec{e}_z})
  + \lambda\hat{P}({\bvec{n}})
  \sum_{m=-\infty}^{\infty}\delta(t - m)
  ,
\end{align}
where $\mu$ and $\lambda$ describe the strength of an applied static
%%{\bf
external field
%%}
and that of the time periodic perturbation, respectively.
%%{\bf
Note that $\lambda$ should be real in physical situation to keep the resultant time evolution unitary.
Later we intentionally extend it into the complex number.
%%}
$\hat{P}(\bvec{v})\equiv (1+\bvec{\pauli}\cdot\bvec{v})/2$ is a projection operator parameterized by a normalized vector $\bvec{v}$, in which $\pauli_j$ ($j=x,y,z$) represents the Pauli matrices, and we set $\hbar=1$%.
%% AT-2009-01-20: a full parameterization of $n$ may simplify the description.
The direction of the perturbed magnetic field is parameterized
as $\bvec{n}=\cos\phi\sin\theta\bvec{e}_x+\sin\phi\sin\theta\bvec{e}_y
+ \cos\theta\bvec{e}_z$.
%%+ \bvec{e}_z\cos\theta$.
%%
%For simplicity, we assume a three-dimensional normalized vector ${\bf n}$ to lie on the $x-z$ plane expressed as $\bvec{n}=\bvec{e}_z\cos\theta +\bvec{e}_x\sin\theta$.
The important physical properties of a periodically driven system are characterized by the so-called Floquet operator, which is a time evolution operator during one period. In particular, an eigenvector of the Floquet operator forms a stationary state, and the corresponding eigenvalue is given as $\exp(-i\gamma)$ with a real $\gamma$ called as a quasienergy. This is nothing but a time domain analogue of Bloch theorem for spatially periodic systems.

For the kicked spin, the Floquet operator is
\begin{align}
  \label{eq:defU}
  \hat{U}(\lambda)
  = e^{-i\mu\hat{P}({\bvec{e}_z})/2}
  e^{-i\lambda\hat{P}({\bvec{n}})}
  e^{-i\mu\hat{P}({\bvec{e}_z})/2}
%   \equiv e^{-i\mu(1+\pauli_z)/4}
%   e^{-i\lambda(1+\bvec{\pauli}\cdot\bvec{n})/2}
%   e^{-i\mu(1+\pauli_z)/4}
  ,
\end{align}
where the fundamental period of the time evolution is chosen as $-1/2 \le t \le 1/2$.
A detailed analysis of $\hat{U}(\lambda)$ is given in Ref.~\cite{Cheon:arXiv:0807.4046}. Since $\hat{U}(\lambda)$ is periodic in $\lambda$ with a period $2\pi$, the parameter space of $\lambda$ is identified with $S^1$. We focus on the closed path $C$ where $\lambda$ is varied from $0$ to $2\pi$ in $S^1$ to examine the quantum holonomies associated with the adiabatic change of the eigenvalues and eigenvectors of $\hat{U}(\lambda)$.

%% AT-2009-02-18: Replace $e^{-i\lambda}$ with $e^{i\lambda}$ to make
%%                the direction of the loop counterclockwise as
%%                $\Re\lambda$ is increased from $0$ to $2\pi$.
\begin{figure}
  \includegraphics[width=5.5cm]{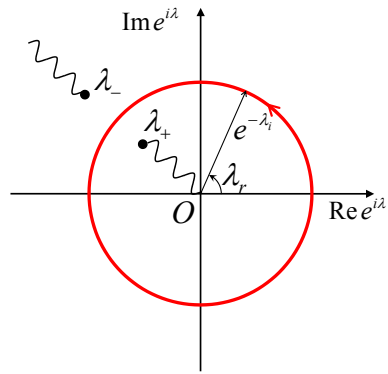}
  \caption{\label{fig1}
    {%\bf
      (Color online) The parameter space extended into complex $\lambda$ for the system described by the unitary matrix (\ref{eq:defU}) by using
      $e^{i\lambda} = e^{-\lambda_i}e^{i\lambda_r}$,
      where $\lambda_r$ and $\lambda_i$ represent the real and imaginary part of $\lambda$, respectively. Two EPs $\lambda_{\pm}\equiv \lambdaR+i\lambdaI$ (black dots) and the corresponding branch cuts of $\Delta(\lambda)$ (curvy lines attached to the dots) are presented, where $\sin\theta\sin({\mu}/{2})>0$ (i.e., $\lambdaI>0$) is assumed. A circle represents the integration contour, in which the angle from the positive $x$ axis and the radius are given as
      $\lambda_r$ and $e^{-\lambda_i}$, respectively.%
    }
  }
\end{figure}

First, we examine the exotic {\em eigenvalue} holonomy. The eigenvalues of $\hat{U}(\lambda)$ are given as
% \endnote{%
%   Since the quasienergies
%   $\left\{\mu + \lambda \pm \Delta(\lambda)\right\}/2$ are
%   defined up to modulus $2\pi$, it is convenient to examine
%   the eigenvalues $z_{\pm}(\lambda)$ instead of the quasienergies here.}:
\begin{align}
  \label{eq:defEigenvalues}
  z_{\pm}(\lambda)
  \equiv
  \exp\left\{-i\left[\mu + \lambda \pm \Delta(\lambda)\right]/2\right\}
  ,
\end{align}
where the gap of the two quasienergies is
\begin{align}
  \label{eq:defDelta}
  \Delta(\lambda)
  &
  \equiv 2\cos^{-1}
  \left[\cos\left(\frac{\lambda-\lambdaR}{2}\right)
    \middle/ \cosh\frac{\lambdaI}{2}\right]
  %%\equiv 2\cos^{-1}\left(- \sin(\lambda/2)\cos\theta\right)
\end{align}
with two real parameters
\begin{subequations}
  \begin{eqnarray}
  \label{eq:defAlpha}
  \lambdaR&\equiv& 2\tan^{-1}\left[-\cos\theta\tan\frac{\mu}{2}\right]
  ,\\
  \label{eq:defBeta}
  \lambdaI&\equiv& 2\tanh^{-1}\left[\sin\theta\sin\frac{\mu}{2}\right]
  .
  \end{eqnarray}
\end{subequations}

The degeneracy then takes place when $\lambda = \lambda_{\pm} + 2\pi k$ is satisfied, where $k$ is an integer and
\begin{align}
  \lambda_{\pm}\equiv\lambdaR \pm i\lambdaI
  .
\end{align}
Note that the degeneracy resides in the complex parameter regime if $\beta \neq 0$. We ignore the case $\beta=0$ at which the problem becomes trivial \cite{endnote:beta neq 0}. Near the degeneracy one approximates Eq.~(\ref{eq:defDelta}) to
\begin{align}
  \Delta \sim 2\sqrt{\pm \frac{i}{2}\tanh\frac{\beta}{2}(\lambda-\lambda_\pm)}
  ,
\end{align}
which implies the degeneracies form square-root branch points of $\Delta(\lambda)$, whose configuration and branch cuts are
depicted in Fig.~\ref{fig1}. This unambiguously reveals the non-Hermitian EPs hidden behind the exotic eigenvalue holonomy. It is well-known that the two eigenvalues are interchanged when the parameters are varied so as to encircle the EP {\em once} \cite{Heiss00,Dembowski:PRE-69-056216}, which is the key feature of the eigenvalue holonomy.

One needs to carefully take into account a closed loop that the parameter is varied.
{%\bf
Naively thinking a semi-infinite rectangular path in $({\rm Re}\lambda,{\rm Im}\lambda)$ containing the real axis from $0$ to $2\pi$ may be considered as a loop. However, the path from $0+i0$ to $0+i\infty$ is exactly identical to that from $2\pi+i0$ to $2\pi+i\infty$ due to the $2\pi$ periodicity of $\lambda$. In fact it is not necessary to depart from the real axis of $\lambda$ to $+i\infty$ in order to encircle the EP.} To understand what is happening here let us redefine the complex parameter space in the polar coordinate,
%% AT-2009-02-18
$e^{i\lambda} = e^{-\lambda_i} e^{i\lambda_r}$,
where $\lambda_r$ and $\lambda_i$ represent the real and imaginary part of $\lambda$, respectively. The EPs and the branch cut may then be represented in Fig.~\ref{fig1}. The exotic holonomy has been found when the {\em real} $\lambda$ is varied from $0$ to $2\pi$, which in fact forms a closed loop encircling the EP, i.e. $\lambda_+$ as is clear from Fig.~\ref{fig1}. It is reemphasized that {\em although a single real parameter $\lambda$ is varied along a closed loop containing no degeneracy on itself this exactly corresponds to encircling a degeneracy hidden in physically inaccessible domain.}

Next, we proceed to the analysis of eigenspace holonomy. To obtain eigenvectors, $\hat{U}(\lambda)$ is casted into a normal form
\begin{align}
  \label{eq:Unormalized}
  \hat{U}(\lambda)
  = \exp\left\{-i\left[\mu + \lambda
      + \Delta(\lambda)\bvec{\pauli}\cdot\bvec{l}(\lambda)
    \right]/2\right\}
    ,
\end{align}
where
%% AT-2009-01-20:
%% (1) Change due to the change of the definition of $\Theta$
%% (2) Add factor $e^{-i\phi}$ for new parametrization of $\bvec{n}$
$
\bvec{l}(\lambda)
=\left(\cos\phi\bvec{e}_x+\sin\phi\bvec{e}_y\right)
\sin\left[2\Theta(\lambda)\right]
+ \cos\left[2\Theta(\lambda)\right]\bvec{e}_z
$
% $
% \bvec{l}(\lambda)
% =\bvec{e}_z\cos\Theta(\lambda)
% +\bvec{e}_x\sin\Theta(\lambda)
% $
and
\begin{align}
  \label{eq:defTheta}
  \Theta(\lambda)&
  =\frac{1}{2}
  \tan^{-1}\frac{\sin\theta}%
  {\sin({\mu}/{2})\cot({\lambda}/{2})
    + \cos\theta\cos({\mu}/{2})}
  .
\end{align}
Let $\ket{\xi_{\pm}(\lambda)}$ be the normalized right eigenvectors of $\hat{U}(\lambda)$ corresponding to $z_{\pm}(\lambda)$, respectively:
\begin{subequations}
  \label{eq:defxipm}
  \begin{eqnarray}
%% AT-2009-01-20: factor $e^{-i\phi}$ for new parametrization of $\bvec{n}$
  \ket{\xi_{+}(\lambda)}
  &=&\cos\Theta(\lambda)\ket{\up}
  +e^{i\phi}\sin\Theta(\lambda)\ket{\down}
  %%+\sin\Theta(\lambda)\ket{\down}
  ,\\
  \ket{\xi_{-}(\lambda)}
  &=&-\sin\Theta(\lambda)\ket{\up}
  +e^{i\phi}\cos\Theta(\lambda)\ket{\down}
  %%+\cos\Theta(\lambda)\ket{\down}
  ,
  \end{eqnarray}
\end{subequations}
%% AT-2009-01-20: fix typo (a redundant comma)
where \mbox{$\ket{\up}$} and \mbox{$\ket{\down}$} are normalized eigenvectors of $\pauli_z$. To incorporate the biorthogonality of the eigenvectors of $\hat{U}(\lambda)$ with a {\em complex} $\lambda$, we denote the eigenvectors of $\{\hat{U}(\lambda)\}^{\dagger}$ as $\ket{\xi^{\rm B}_{\pm}(\lambda)}$, on which we impose the biorthonormal relation $\bracket{\xi^{\rm B}_{m}(\lambda)}{\xi_{n}(\lambda)} = \delta_{mn}$%
~\cite{Garrison:PLA}.

Following the prescription introduced in Ref.~\cite{Cheon:arXiv:0807.4046}
%% AT-2009-01-20: Change the name of $A$
a non-Abelian gauge connection
%%the non-Abelian gauge potential
%%
is
%%{\bf
given
%%}
as
\begin{equation}
  \label{eq:defAFexample}
  \AF_{mn}(\lambda) = i \left<\xi^{\rm B}_m\right|\frac{\partial}{\partial \lambda}\left|\xi_n\right>.
\end{equation}
%\end{align}
We denote the diagonal part of $\AF(\lambda)$ as $\AD(\lambda)$, whose elements provide the Mead-Truhlar-Berry's adiabatic gauge
%% AT-2009-01-20: Change the name of $\AD$
connections%
%%potentials
~\cite{Mead:JCP-70-2284,Berry:PRSLA-392-45}. With an arbitrary choice of the phases of $\ket{\xi_\pm}$, the gauge covariant expression of the holonomy matrix for a given closed path $C$ reads
\begin{align}
  \label{eq:defHolonomyMatrix}
  &
  M(C)
  \nonumber\\ &
  =\AntiTexp\left(-i\int_C \AF(\lambda)d\lambda\right)
  \Texp\left(i\int_C \AD(\lambda)d\lambda\right)
  ,
\end{align}
where $\Texp$ and $\AntiTexp$ indicate the path-ordered and the anti-ordered
%% AT-2009-01-20: I believe 'ordered exponential' is not a jargon
%% (e.g. http://scholar.google.com/scholar?q="path-ordered+exponential" )
exponentials, respectively~\cite{Cheon:arXiv:0807.4046}. The first factor in the right hand side describes the transformation of the basis induced by the transport along $C$ and is expected to be a permutation matrix that reflects the key feature of the eigenspace holonomy. Both the overall phase of the first factor and the second factor describe the Berry phase. When we choose the gauge such that satisfies the parallel transport condition, $\AD(\lambda)=0$, the second factor becomes the identity.

%% AT-2009-01-20: Change the name of $\AF$
The gauge connection for the kicked spin-$1/2$ is obtained from Eq.~(\ref{eq:defAFexample}) as
\begin{align}
  \AF(\lambda)
  =
  \begin{bmatrix}
    0&-i\\ i& 0
  \end{bmatrix}
  \pdfrac{\Theta}{\lambda}
  ,
\end{align}
where the parallel transport condition $\AD(\lambda)=0$ is satisfied by choosing the phases of $\ket{\xi_\pm}$. $M(C)$ is then directly acquired from Eq.~(\ref{eq:defHolonomyMatrix});
\begin{align}
  \label{eq:evalH}
  M(C)
  =
    \begin{bmatrix}
      \cos\left(\eta(C)\right)&-\sin\left(\eta(C)\right)\\
      \sin\left(\eta(C)\right)& \cos\left(\eta(C)\right)
    \end{bmatrix}
  ,
\end{align}
where
\begin{equation}
  \label{eq:defEta}
  \eta(C)
  =
  \oint_{C}\pdfrac{\Theta}{\lambda}d\lambda
  .
\end{equation}

It is easily shown that $\eta(C)=\sgn(\beta)\pi/2$ from the residue of the pole of the integrand which is located at the EP enclosed by the loop in Fig.~\ref{fig1}, where $\sgn(x)=1$ for $x>0$ otherwise $-1$ (Recall that $\beta \neq 0$). It immediately gives
\begin{align}
  \label{eq:evalHResult}
  M(C)
  =
  \sgn(\beta)
  \begin{bmatrix}
    0& -1\\ 1& 0
  \end{bmatrix}
  ,
\end{align}
which is a permutation matrix, as we expected, except the minus sign. The meaning of the minus sign becomes clear if $\lambda$ is varied along the loop {\em twice}. According to the geometry of the Riemann surface around the EP the exact initial condition can be recovered only by encircling the EP twice. The holonomy matrix is then given as
\begin{align}
%% AT-2009-01-20: Fix for duplication of the label
  \label{eq:evalHResultC2}
  %%\label{eq:evalHResult}
  M(C^2)
  =
  \begin{bmatrix}
    -1 & 0\\ 0 & -1
  \end{bmatrix}
  .
\end{align}
Here we obtain the parallel transport with a phase change, namely $\pi$, which is nothing but the Berry phase. It implies the minus sign in
%% AT-2009-01-20: Fix for duplication of the label
Eq.~(\ref{eq:evalHResult}) also comes from the Berry phase.

%% AT-2009-02-18: Now we have three remarks...
%Several remarks are in order.
%%
In mathematical sense, one may conclude that the exotic holonomies originate simply from the EPs, i.e., the square root branch points of eigenvalues in the complex parameter space. Note that even a simple one-dimensional nonlinear oscillator may possess an infinite number of branch points~\cite{Bender:PR-184-1231}. In this sense, one can say that the exotic holonomies are ubiquitous and hidden everywhere.
However, these are only suprficially true.
Although it is worth elaborating the mathematical analogy further,
the physical significance of the exotic holonomies and the EPs
must be distinguished deliberately.
To clarify it, three important implications of our finding are discussed.
%%However, this is not true.
%%In fact our finding has three important implications.

First, the usual gauge theory of holonomies is also applicable to the holonomy for an open path in the sense that one should encircle the EP twice to complete the path due to the geometry of the Riemann surface. This feature has been overlooked in the conventional treatise of the phase holonomy although it has already been examined experimentally in the context of open quantum systems~\cite{Dembowski:PRL-86-787}.

{%\bf
%In mathematical sense, one may conclude that the exotic holonomies originate simply from the EPs, i.e., the square root branch points of eigenvalues in the complex parameter space. However, this is not true. In fact our finding has three important implications. First, the usual gauge theory of holonomies is also applicable to the holonomy for an open path in the sense that one should encircle the EP twice to complete the path due to the geometry of the Riemann surface. This feature has been overlooked in the conventional treatise of the phase holonomy although it has already been examined experimentally in the context of open quantum systems~\cite{Dembowski:PRL-86-787}.

Second, in the physical point of view the degeneracy generating the exotic holonomy is distinguished from the EP. The EP has been discussed in the context of an open quantum system \cite{Muller08}, in which an effective model Hamiltonian is no longer Hermitian and the corresponding eigenvalues are complex numbers. Their imaginary part has a definite physical meaning, namely the decay time of the state. The extension of the eigenvalues into the complex domain thus intrinsically occurs. In our case, instead, we do not deal with any model Hamiltonian including a non-Hermitian part, so that the eigenvalues remain in the unit circle to keep the unitarity of the whole adiabatic time evolution. The degeneracy is hidden in the unphysical domain described by the parameter space intentionally extended into the complex number. Thus its physical implication is quite different from that of the usual EP.

%% AT-2009-02-18
Finally, we point out the relationship between the exotic holonomies and nonadiabatic (or Landau-Zener) transitions~\cite{LandauLifshitzBranchPoint}
in view of quantum state manipulations.
Both of them induce ``excitations'' of quantum state by the parametric evolution, and are mathematically governed by hidden degeneracies
in complex space.
%Finally, the exotic holonomy seems to have something to do with a non-adiabatic (or Landau-Zener) transition transition~\cite{LandauLifshitzBranchPoint} since both of them induces ``excitations'' of quantum state by the parametric evolution whatever adiabatically or non-adiabatically they are.
%% AT-2009-02-18: slightly simplified
As for the non-adiabatic transitions, the excitation of quantum state occurs
with only exponentially small probability along an almost adiabatic process.
%As for the non-adiabatic transitions, the excitation of quantum state occurs due to non-adiabatic perturbation so that the transition rate becomes exponentially small for adiabatic process.
The exotic holonomy, however, allows us to achieve almost perfect excitation.
%In the exotic holonomy, however, almost perfect excitation can be achieved not by perturbing the states but rather by directly exchanging the two interacting eigenstates.
It may be exploited as a new way of exciting a system with branch points by adiabatic process.
%%
%% AT-2009-02-18: Maybe the above and the below are somewhat different subject
%%                So moved the following to the first paragraph of discussion.
%Note that even a simple one-dimensional nonlinear oscillator may possess an infinite number of branch points~\cite{Bender:PR-184-1231}. In this sense, one can say that the exotic holonomies are ubiquitous and hidden everywhere.
}

%%\paragraph{Summary}

In summary we have investigated the exotic holonomies focusing on its origin using the periodically kicked spin-$\frac{1}{2}$ system. All the non-trivial characteristics of the exotic holonomies is ascribed to existence of the degeneracy hidden in the complex parameter space.
Our finding delivers an important message that one might encounter the unexpected non-trivial holonomies originated from the degeneracy hidden in the complex parameter space which is usually ignored.
{%\bf
  It might play an important role in many areas of physics.}

%% seems to be trivial...
%% - (?) Physical reality of branch point
%%  A bound state "feels" the branch point through adiabatic time evolution.

\begin{acknowledgments}

This work was supported by Korea Research Foundation Grant (KRF-2006-005-J02804 and KRF-2008-314-C00144), and also by
the Grant-in-Aid for Scientific Research of  Ministry of Education,
Culture, Sports, Science and Technology, Japan (Grant Number18540384).

\end{acknowledgments}

%%%
%%% for preparation, I prefer to use BibTeX
%%%
%%%\bibliography{holonomy}

%%%
%%% for submission, we need to embed the .bbl file here
%%%

%%%
%%% end of .bbl
%%%

%\onecolumngrid\newpage
%\noindent{\tt MEMO (to be deleted)}
%\input{memo}

\end{document}